\documentclass{aa}  
\usepackage{graphicx}
\usepackage[varg]{txfonts}
\usepackage{natbib}
\usepackage{amssymb}
\usepackage{booktabs,siunitx}
\usepackage{dcolumn}
\bibpunct{(}{)}{;}{a}{}{,}

\def\tex {\ifmmode{{T}_{\rm ex}}\else{$T_{\rm ex}$}\fi}
\def\tmb {\ifmmode{{T}_{\rm mb}}\else{$T_{\rm mb}$}\fi}
\def\ci     {\ifmmode{{\rm C}{\rm \small I}}\else{C\ts {\scriptsize I}}\fi}
\def\hi     {\ifmmode{{\rm H}{\rm \small I}}\else{H\ts {\scriptsize I}}\fi}
\def\hh     {\ifmmode{{\rm H}_2}\else{H$_2$}\fi}

\def\ts     {\thinspace}
\def\kms    {\ifmmode{{\rm \ts km\ts s}^{-1}}\else{\ts km\ts s$^{-1}$}\fi}
\def\msol   {\ifmmode{{\rm M}_{\odot}}\else{M$_{\odot}$}\fi}
\def\lsol   {\ifmmode{{\rm L}_{\odot}}\else{L$_{\odot}$}\fi}
\def\zsol   {\ifmmode{{\rm Z}_{\odot}}\else{Z$_{\odot}$}\fi}

\begin{document}

   \title{Ram Pressure Stripping in the Virgo Cluster}

   \author{C. Verdugo\inst{1},  F. Combes\inst{1,2}, K. Dasyra\inst{1,3}, P. Salom\'e \inst{1}
          \and
          J. Braine\inst{4}
          }

\offprints{C. Verdugo}
\institute{Observatoire de Paris, LERMA (CNRS:UMR8112), 61 Av. de l'Observatoire, F-75014, Paris, France
\email{celia.verdugo@obspm.fr}
 \and
Coll\`ege de France, 11 Place Marcelin Berthelot, 75005 Paris, France
 \and
Department of Astrophysics, Astronomy \& Mechanics, Faculty of Physics, University
of Athens, Panepistimiopolis Zografos 15784, Greece
 \and
Univ. Bordeaux, Laboratoire d'Astrophysique de Bordeaux, (CNRS:UMR5804) 33270 Floirac, France
              }

   \date{Received 2015; accepted 2015}

   \titlerunning{CO in the Virgo Cluster}
   \authorrunning{C. Verdugo et al.}


   \abstract{Gas can be violently stripped from their galaxy disks in rich clusters,
and be dispersed over 100kpc-scale tails or plumes. Young stars have been observed in these tails,
suggesting they are formed in situ. This will contribute to the intracluster light,
in addition to tidal stripping of old stars. We want to quantify the efficiency 
of intracluster star formation.
We present CO(1--0) and CO(2--1) observations, made with the IRAM-30m telescope, towards
the ram-pressure stripped tail northeast of NGC4388 in Virgo. HII regions found
all along the tails, together with dust patches have been targeted. We detect 
molecular gas in 4 positions along the tail, with masses between  7x10$^5$ to 2x10$^6$ M$_\odot$.
Given the large distance from the NGC 4388 galaxy, the molecular clouds must have
formed in situ, from the HI gas plume. We compute the relation between
surface densities of star formation and molecular gas in these regions,
and find that the star formation has very low efficiency. 
The corresponding depletion time of the molecular gas can be up to 500 Gyr and more. 
Since this value exceeds a by far Hubble time, this gas will not be converted into stars, and will stay 
in a gaseous phase to join the intracluster medium.}


   \maketitle
%

\section{Introduction}

In overdense cluster environments, galaxies are significantly transformed, through different
tidal interactions, like the ones due to  other galaxies, the cluster as a whole
(e.g. \citealt{merritt1984}, \citealt{tonnesen2007}),
and the ones with the intra-cluster medium (ICM), which strips them from their gas content.
This ram-pressure stripping (RPS) process has been described by \cite{gunn1972} and simulated 
by many groups (\citealt{quilis2000}, \citealt{vollmer2001}, \citealt{roediger2005}; \citealt{jachym2007}).
Evidence of stripping has been observed in many cases (\citealt{kenney2004}; \citealt{chung2007};
\citealt{sun2007}, \citealt{vollmer2008}). RPS and/or tidal interactions can disperse the interstellar gas (ISM) of galaxies
at large distance, up to 100kpc scales, as shown by the spectacular tail of ionized
gas in Virgo \citep{kenney2008}.

What is the fate of the stripped gas? According to the time-scale of the ejection, the relative velocity
of the ICM-ISM interaction, and the environment, it could
be first seen as neutral atomic gas (\citealt{chung2009}, \citealt{scott2012}, \citealt{serra2013}),
then ionized gas detected in H$\alpha$ (\citealt{gavazzi2001},
\citealt{cortese2007}, \citealt{yagi2007}, \citealt{zhang2013}), and is finally heated to X-ray gas temperatures
(e.g. \citealt{machacek2005}, \citealt{sun2010}). In rarer cases, it can be seen as 
dense and cold molecular gas, detected as carbon monoxyde (CO) emission (\citealt{vollmer2005}, \citealt{dasyra2012},
\citealt{jachym2014}).  The presence of these dense molecular clumps
might appear surprising, since the RPS should not be able to drag them
out of their galaxy disks (\citealt{nulsen1982}; \citealt{kenney1989}). However they could reform
quickly enough in the tail. The survival of these clouds in the hostile ICM
environment, with temperature 10$^7$ K and destructive X-rays (e.g., \citealt{machacek2004}; 
\citealt{fabian2006};
\citealt{tamura2009})  is a puzzle, unless they are self-shielded
(e.g. \citealt{dasyra2012}, \citealt{jachym2014}).  The presence of cold molecular gas
is also observed in rich galaxy clusters, with cool cores. Here also a multi-phase gas
has been detected, in CO, H$\alpha$, X-rays and also the strongest atomic cooling lines
\citep{edge2010}.  Ionized gas, together with
warm atomic and molecular gas and cold molecular gas clouds 
coexist in spatially resolved filaments around the brightest cluster galaxy, such as in the spectacular
prototype Perseus A (\citealt{conselice2001}; \citealt{salome2006,salome2011}; \citealt{lim2012}).

The survival of molecular clouds was also observed by \cite{braine2000} in several tidal
tails, and in particular in the interacting system 
Arp 105 (dubbed the Guitar), embedded in the X-ray emitting medium of the Abell 1185 
cluster \citep{mahdavi1996}. Again, the formation in
situ of the molecular clouds is favored \citep{braine2000}. In the Stephan's Quintet
compact group, where X-ray gas and star formation have been observed in between galaxies
 \citep{osullivan2009},
the shock has been so violent (1000 km/s) that H$_2$ molecules are formed and provide
the best cooling agent, through mid-infrared radiation \citep{cluver2010}. In this shock, multi-phases
of gas coexist, from cold dense molecular gas to X-ray gas.

Does this gas form stars? In usual conditions, inside galaxy disks, the star formation
is observed to  depend on the amount of molecular gas present (e.g. \citealt{bigiel2008};
\citealt{leroy2013}). A  Schmidt-Kennicutt (S-K) relation is observed, roughly linear, between 
the surface densities of star formation and molecular gas, leading to a depletion time-scale
($\tau_{\rm dep}=\Sigma_{\rm  gas}/\Sigma_{\rm SFR}$)
of 2 Gyr. But this relation does not apply in  particular regions or circumstances,
such as galaxy centers \citep{casasola2015}, outer parts of galaxies and 
extended UV disks \citep{dessauges2014}, or low surface brightness galaxies
\citep{boissier2008}. 
Little is known on star formation in gas clouds stripped from galaxies in rich clusters.
\cite{boissier2012} have put constraints on this process, concluding to a very
low star formation efficiency, lower by an order of magnitude than what is usual
in galaxy disks, and even lower than outer parts of galaxies or in low surface 
brightness galaxies.
 It is interesting to better constrain this efficiency, given the large amount of intracluster light (ICL)
observed today (e.g. \citealt{feldmeier2002}, \citealt{mihos2005}).
These stars could come from tidal stripping of old stars formed in 
galaxy disks, or also a large fraction could have formed in situ, from ram-pressure
stripped gas. More intracluster star formation could have formed in the past \citep{demaio2015}. 
The origin of the ICL could bring insight on the relative role
of galaxy interactions during the cluster formation, or cluster processing after relaxation. 

\subsection{The tail northeast of NGC4388}

 One of the environments to probe the survival of molecular gas and the efficiency of star formation
under extreme ram-pressure conditions
is the RPS tail north of NGC 4388
in the Virgo cluster south of M86, where X-ray gas has been mapped
\citep{iwasawa2003} and young stars have been found \citep{yagi2013}. 
It is located at about 400 kpc in projection from the cluster center M87. NGC 4388 is
moving at a relative velocity redshifted by 1500km/s with respect to M87,
and more than 2800 km/s with respect to the M86 group. This strong
velocity may explain the violent RPS, the high HI deficiency of NGC 4388 \citep{cayatte1990}
and the large ($\sim$ 35kpc) emission-line region
found by \cite{yoshida2002}, northeast of the galaxy. The ionized gas has a mass of 
10$^5$ M$_\odot$, and is partly excited by the ionizing radiation of the Seyfert 2 nucleus
in NGC 4388.  The RPS plume is even more extended in HI \citep{oosterloo2005},
up to 110 kpc, with a mass of  3.4 x 10$^8$ M$_\odot$. \cite{gu2013}
have found neutral gas in absorption in X-ray, with column densities 
2-3$\times$10$^{20}$ cm$^{-2}$, revealing that the RPS tail is in front of M86. The high ratio between hot and 
cold gas in the clouds means that significant evaporation has proceeded.
\cite{yagi2013} find star-forming regions in the plume at 35 and 66 kpc
from NGC 4388, with solar metallicity and age 6 Myrs. Since these stars are younger
than the RPS event, they must have formed in situ.

In the present paper we present CO detections in the ram-pressure stripped gas
northeast of NGC 4388.
In a previous paper, we have already found molecular gas in a ionized gas tail
south of M86 \citep{dasyra2012}, and discussed its survival conditions.
We  here study the link between new stars formed and molecular gas, to
derive the star formation efficiency. In the RPS plume, a
significant fraction of the H$\alpha$ emission could
originate from the ionized gas in the outer layers of molecular clouds \citep{ferland2009}. 
This makes the H$\alpha$ lumps good tracers of star formation in an RPS tail, to probe
the efficiency of the process of formation of intracluster stars.
 Section 2 presents the IRAM-30m observations, Section 3 the results obtained,
which are discussed in Section 4.

In the following, we assume a distance of 17.5 Mpc to the Virgo cluster (Mei et al. 2007).


\section{Observations and data reduction}

CO observations along the  HI plume \citep{oosterloo2005} connecting NGC4388 and M86 were done with the IRAM
30-mt telescope at Pico Veleta, Spain, in two separate runs. 
The first run was part of the project 195-13, with 28 hours of observation, and took place between the 5th and 8th of December 2013,
with excellent weather conditions ($\tau<0.1$ and a pwv between 0.1 and 3mm).
The second run was project 075-14, with 47 hours of observations between June 25th-30th 2014, 
and had poor to average weather conditions ($\tau$ between 0.2 and 0.6 and a pwv between 3 and 10 mm).

All observations were done with the EMIR receiver in the E0/E2 configuration, allowing us to observe simultaneously 
CO(1--0) and CO(2--1) at 115.271 and 230.538 GHz respectively.
The telescope half-power beam widths at these frequencies are 22\arcsec~and 11\arcsec~respectively.
The observing strategy consisted in single ON+OFF pointings per each target, with wobbler switching.

\begin{figure*}
\centering
\includegraphics[width=0.85\textwidth]{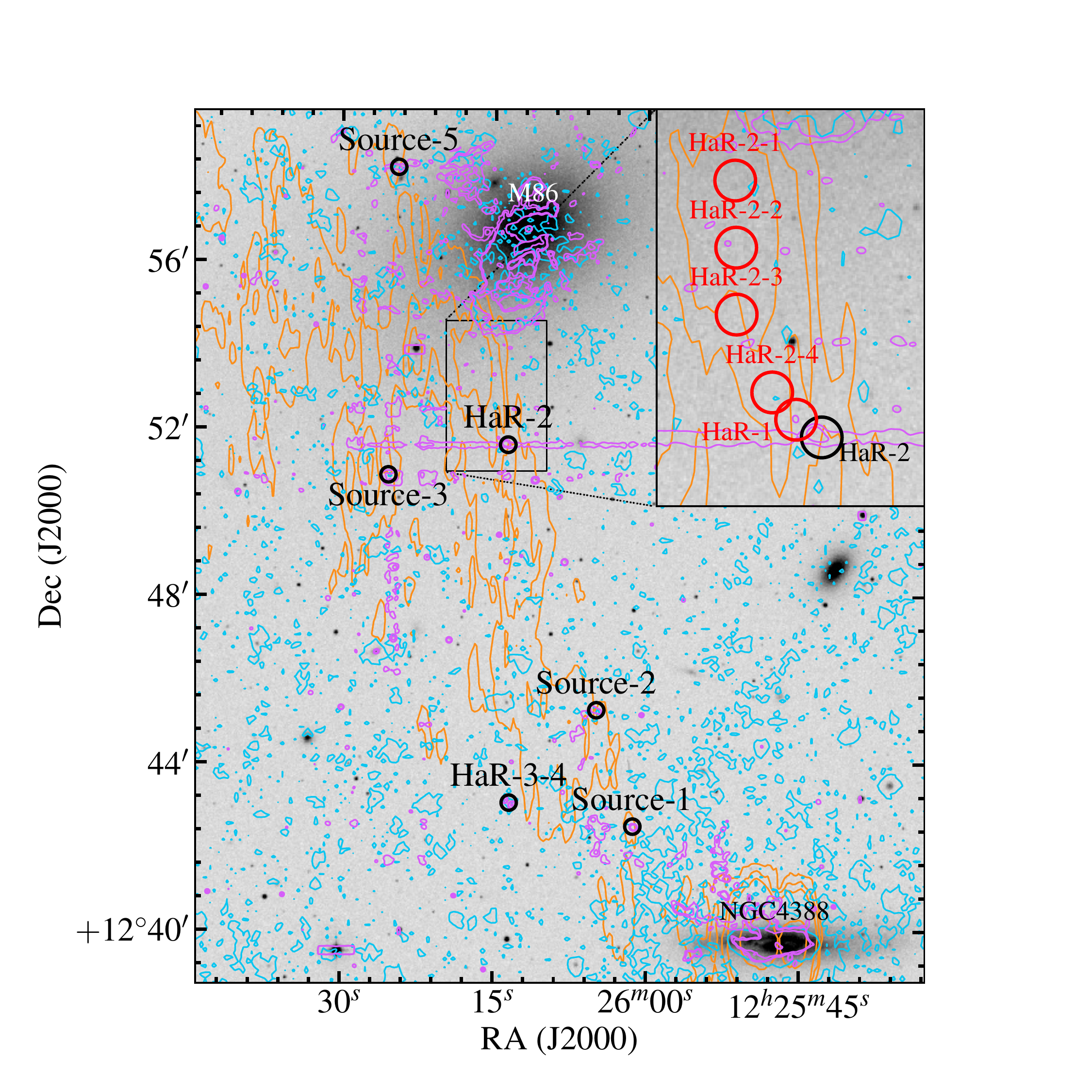}
\caption{Targets observed in the Virgo Cluster. Orange: HI contours levels at 1, 5, 10 and 50  
$\times 10^{19}cm^{-2}$ from \cite{oosterloo2005},  showing the plume of atomic gas stripped from NGC4388. 
Purple: H$\alpha$  contours levels at  
5, 11 and 50 $e^{-}/sec$ \citep{kenney2008}. Cyan: 250$\mu$m contour levels at 0.01 and 0.1 Jy/beam \citep{davies2012}.
In zoomed regions: targets selected close to HaR-2 for run 075-14 (red circles).
Circles enclosing targets are 22\arcsec~width, as the CO(1--0) HPBW.}
\label{fig:virgo_map}%
\end{figure*}

Targets along the HI plume were selected for having a match of HI (using the N$_{\rm HI}$ map from \citealt{oosterloo2005}), 
H$\alpha$ (with data from \citealt{kenney2008} and \citealt{yagi2013}) and 250$\mu$ emission (HERSCHEL SPIRE data from 
\citealt{davies2012}). See Figure \ref{fig:virgo_map}. 
With this criteria 6 targets where selected for the first run 195-13, and are listed in Table \ref{tab:targets} 
(first 6 rows).
As a result of this run, only two sources showed CO detection: Source-1 and HaR-2.
Since HaR-2 is of particular interest for being so far away from both galaxies and for having a strong H$\alpha$
detection \citep{yagi2013}, it was chosen as a central target around which other 5 extra targets were selected for the second run 075-14 (second half of Table \ref{fig:virgo_map}), following the path of an HI peak (Figure \ref{fig:virgo_map} top right box)

\begin{table}[h]
\caption{Targets and Observations \label{tab:targets}}
\centering
\begin{tabular}{ccccc}
\hline\hline
Source & RA(J2000) & DEC(J2000) & hel. vel. & ON+OFF \\
 & [$^{hr}$:$^{m}$:$^s$] & [$^{deg}$:$^m$:$^s$] & [km/s] &[hrs] \\
\hline
Source-1 & 12:26:01.3 & 12:42:30.1 & +2500 &  4.2\\
Source-2 & 12:26:04.9 & 12:45:16.7 & +2500 &  2.9\\
Source-3 & 12:26:25.4 & 12:50:53.6 & +2200 & 3.1\\
Source-5 & 12:26:24.5 & 12:58:14.6 & +2000 &  3.9\\
HaR-2 & 12:26:13.7 & 12:51:36.9 & +2230 &  7.2\\ 
HaR-3-4 & 12:26:13.5& 12:43:03.7 & +2500 & 3.8 \\
\hline
HaR-2-1 & 12:26:16.9 &12:53:55.6 & +2230 & 4.5\\
HaR-2-2 & 12:26:16.9 & 12:53:19.3  & +2230 & 5.6\\
HaR-2-3 & 12:26:16.9 &12:52:43.1 & +2230 & 6.2\\
HaR-2-4 & 12:26:15.5 & 12:52:01.1 & +2230 & 8.0\\
HaR-1 & 12:26:14.6 & 12:51:46.4 & +2230 & 3.8\\
NGC4388 & 12:25:46.6 & 12:39:44.0 & +2550 & 1.2\\
\hline
\end{tabular}
\tablefoot{Sources in the first 6 rows correspond to run 195-13, and last 6 rows to run 075-14 (although source HaR-2 was observed in both runs).
NGC4388 was observed for calibration purposes.
Heliocentric velocities are referencial, taken from \cite{oosterloo2005}  along the HI plume.}
\end{table}

Concerning the spectral resolution of our data, during the observations both FTS and WILMA backends were used, simultaneously.
The FTS backend has a spectral resolution of 195 kHz and a bandwidth of 32 GHz including both polarizations.
At 115 GHz these values correspond to 0.5 and 83200 km/s, and at 230 GHz to 0.25 and 41600 km/s.
As for the WILMA backend, we obtained a spectral resolution of 2 MHz and a bandwidth of 16 GHz.
At 115 GHz this translates to 5.2 and 41600 km/s, and to 2.6 and 20800 km/s at 230 GHz.

\begin{figure*}
\centering
\includegraphics[width=0.8\textwidth]{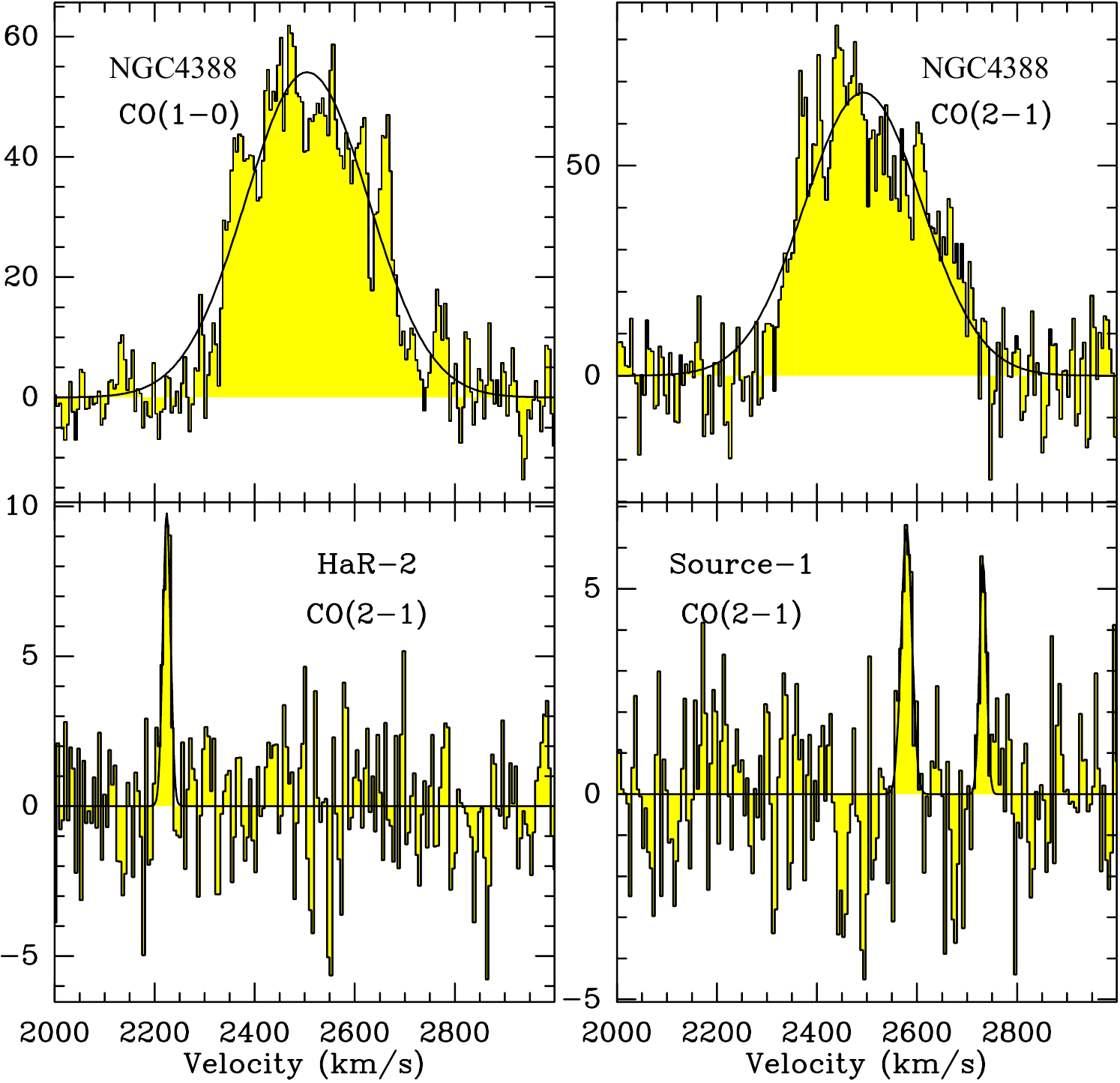}
\caption{Top: Spectra taken at the center of NGC4388, observed for calibration purposes.
Bottom: Spectra from observing run 195-13 that showed CO emission. 
Spectra information and line parameters are presented in 
Table \ref{tab:line_parameters}. The temperature scale corresponds to main beam temperature in mK.}
\label{fig:spectra-1}
\end{figure*}

\begin{figure*}
\centering
\includegraphics[width=0.8\textwidth]{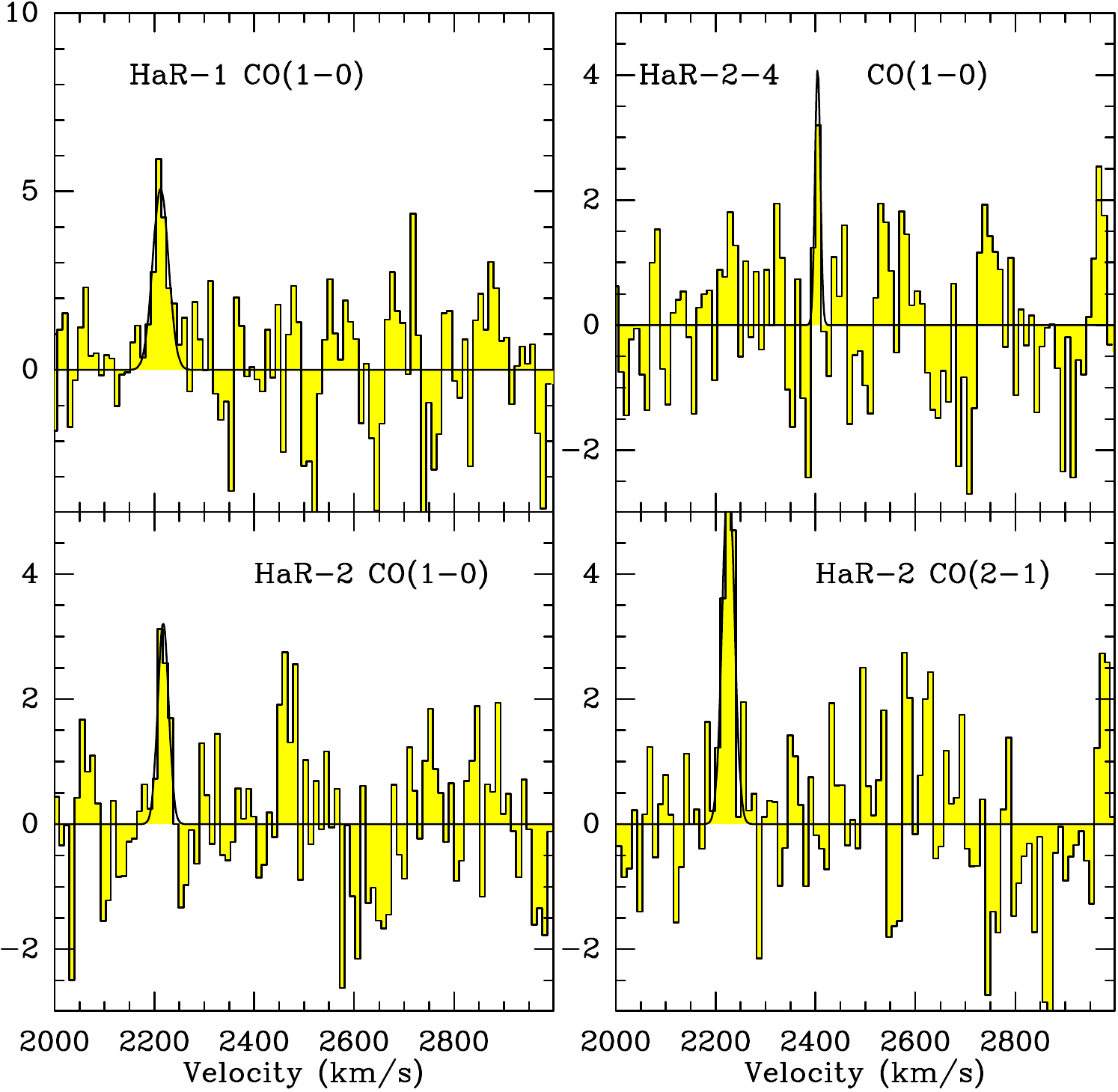}
\caption{Final CO spectra for HaR-2, HaR-1 and HaR-2-4 from run 075-14. Only HaR-2 in CO(2--1) has data combined from both runs. All spectra have both polarizations combined. Spectra information and line parameters are presented in Table \ref{tab:line_parameters}. The temperature scale corresponds to main beam temperature in mK.}
\label{fig:spectra_2}
\end{figure*}

The data were reduced using the CLASS software from the GILDAS package.
First, a careful inspection of all scans was done, to remove the bad ones.
The approved scans of the same source, CO line and backend were averaged with a normal time weighting, to obtain one  spectrum.
Then, each spectrum was inspected individually, and in both of its polarizations, to identify a possible CO emission line.
If a detection was found in the spectra of both backends, the best spectrum was chosen (in terms of 
spectral resolution and S/N) as the final one.
The selected spectra with CO emission are presented in Figures \ref{fig:spectra-1} 
and \ref{fig:spectra_2}, which contain both polarizations, horizontal and vertical, combined.

Baselines were subtracted with polynomials of order 0 and 1 depending on the source, and antenna temperatures were corrected
by the telescope beam and forward efficiencies\footnote{http://www.iram.es/IRAMES/mainWiki/Iram30mEfficiencies}
to obtain main beam temperatures.
Spectra were smoothed with the hanning method to degrade the velocity resolution until obtaining a value
no greater than 1/3 of the FWHM line.

Finally, a simple gaussian line was fitted to the line candidate. 
The CLASS fit return the velocity position of the line, its FWHM, the peak temperature and the integrated line intensity.

Such spectra and fitting results for sources with CO detection are presented in Figures \ref{fig:spectra-1} (from the first run)
and \ref{fig:spectra_2} (from second run) and then in Table \ref{tab:line_parameters}.
For the rest of the sources, with no visible CO detections, 3$\sigma$ upper limits for I$_{\rm CO}$ where calculated
from their rms values, assuming a $\Delta$v = 30 km/s . Such limits are presented in Table \ref{tab:upperlimits}.

 \section{Results}

After selecting the final spectra for every source with a visible detection and fitting their gaussian profiles, line parameters are calculated and
presented in Table \ref{tab:line_parameters}. 
These CO detections present velocity centroids in the range of $\sim$2200-2500 km/s, and are quite consistent with the HI velocities
from Table \ref{tab:targets}, taken from \citealt{oosterloo2005} (their Fig. 2).
H$_2$ masses were derived from the CO(1--0) line intensity, using a Galactic CO 
conversion factor of $2.0\times10^{20}$[cm$^{-2}$(K km/s)$^{-1}$] and a correction factor of 1.36 to account for heavy elements:

\begin{center}
\begin{equation}
\rm  M_{H_2}[M_{\odot}]=4.4\pi R^2[pc]~ I_{CO(1-0)}[K~km/s]
\label{eq:M_H2}
 \end{equation}
\end{center}

where the source's radius R corresponds to the CO(1--0) beamsize radius at the distance of the source (17.5 Mpc from \citealt{mei2007}).

\begin{table*}
\caption{CO detections and their line parameters.\label{tab:line_parameters}}
\centering
\begin{tabular}{c c c c D{,}{\pm}{-1} D{,}{\pm}{-1} c c c D{,}{\pm}{-1} c}
\hline\hline
Source & line & backend & int. time & \multicolumn{1}{c}{v$_0$} & \multicolumn{1}{c}{FWHM} & resol. & T$_{\rm mb}$ & rms &\multicolumn{1}{c}{ I$_{\rm CO}$} & M$_{\rm H_2}$ \\
 & & &[min]& \multicolumn{1}{c}{[km/s]} & \multicolumn{1}{c}{[km/s]} &[km/s] & [mK] & [mK] & \multicolumn{1}{c}{[mK km/s]}  & [10$^6$M$_{\odot}$]\\
\hline
HaR-2 &	CO(2--1) &	FTS &	268 &	2224,2 &	18,3	 & 4 & 	9.8 &	2.0 &	186,29 & \\		
Source-1(1) &	CO(2--1) &	WILMA &	232 &	2579,2 &	25,5 &	5 &	6.4 &	1.7 &	173,29 & \\		
Source-1(2)	& CO(2--1) &	 WILMA & 	232 &	2732,2 &	16,5 &	5 &	5.6 &	1.7 &	99,25	& \\	
\hline
HaR-2 &	CO(1--0) &	WILMA &	318 &	 2218,4 &	 26,7 &	10 & 	3.2 &	1.5 & 	88,25 & 1.1 \\
&	CO(2--1) &	WILMA &	586 &	2226,2 & 24,6 & 10 & 	6.3 &	1.6 &	163,29 & \\		
HaR-1 &	CO(1--0) &	WILMA &	229 &	2212,4 & 35,13 &	10 & 	5.2 &	2.3 &	199,54 & 2.4 \\
HaR-2-4 &	CO(1--0) &	WILMA &	665 &	2404,2 &	12,3 &	10 & 	4.0 &	1.4 &	54,16 & 0.7 \\
& \\		
NGC4388 &	CO(1--0) &	WILMA &	87 &  2505,2 &	289,5 &	5 &	54.0 &	4.8 &	17000,276 &	205 \\
 &	CO(2--1) &	WILMA &	82 &	 2493,4 & 275,7 &	5 & 67 &  8.9 & 20000,490	& \\	

\hline
\end{tabular}
\tablefoot{First three rows correspond to observing run 195-13. Both components shown in Figure \ref{fig:spectra-1} for Source-1 are listed.
H$_2$ masses were calculated as M$_{\rm H_2}$[M$_{\odot}$]=4.4$\pi$R$^2$[pc] I$_{\rm CO}$[K km/s], which uses 
a Galactic CO conversion factor of 2.0$\times$10$^{20}$ [cm$^{-2}$(K km/s)$^{-1}$] and a correction factor of 1.36 to account for heavy elements.
A distance to the Virgo Cluster of 17.5 Mpc \citep{mei2007} was used to calculate the CO(1--0) beam radius as source radius.
HaR-2 includes data taken in the first run only for CO(1--0), and data combined from both runs for CO(2--1).}
\end{table*}

\begin{table*}
\caption{CO(1--0)  upper limits at 3$\sigma$ for sources with no detection.\label{tab:upperlimits}}
\centering
\small
\begin{tabular}{lccccc}
\hline \hline
Source & int. time & rms & I$_{\rm CO}$\tablefootmark{a} & M$_{\rm H_2}$ & $\Sigma_{\rm H2}$ \\
 & [min] & mK & [K km/s]& [10$^6$M$_{\odot}$] & [M$_{\odot}$pc$^{-2}$]\\
 \hline
Source-2 & 173 & 1.2 & <0.11 & <1.21 & <0.48\\
Source-3 & 186 & 1.0 & <0.09 &<0.99 & <0.40\\
Source-5 & 231 & 1.2 & <0.11 & <1.21 & <0.48\\
HaR-3-4 & 229 & 1.0 & <0.09 & < 0.99 & <0.40\\
HaR-2-1 & 269 & 1.1 & <0.10 & <1.10 & < 0.44\\
HaR-2-2 & 334 & 1.1 & <0.10 & <1.10 & <0.44\\
HaR-2-3 & 374 & 0.6 & <0.05 & <0.55 & <0.22\\
\hline
\end{tabular}
\tablefoot{
\tablefoottext{a}{We assumed a $\Delta$v of 30 km/s.}}
\end{table*}
\subsection{Star formation efficiency}

To estimate how fast is the gas being transformed into stars, we compare the SFR surface density versus the gas surface density
in a  S-K relation, to understand the efficiency of this star formation process. 
Since these are low gas density regions, we can expect a large amount of gas in atomic phase, greater than in molecular phase.
Therefore both components, atomic and molecular, need to be taken into account when estimating a total amount of gas to be converted into stars.

Molecular gas can be directly estimated from the CO(1-0) line intensity, obtained from our observations. 
If we take Eq. \ref{eq:M_H2}, and we divide it by the source area ($\pi$R$^2$),
we obtain the H$_2$ surface density:
 
\begin{center}
\begin{equation}
\rm \Sigma_{H_2}[M_{\odot}pc^{-2}] = 4.4 I_{CO(1-0)}[K~km~s^{-1}]
\label{eq:Sigma_H2}
\end{equation}
\end{center}

These values are listed in column 2 of Table \ref{tab:S-K_values} for the sources with CO(1--0) detections, including  upper limits for the sources with no CO detection using Table \ref{tab:upperlimits}.

For the atomic gas component, we estimated the amount of HI from the HI column density map of the NGC4388 plume from \cite{oosterloo2005}.
The atomic gas mass is derived from the 
 integrated amount of N$\rm _H$ inside the source solid angle:
\begin{center}
\begin{equation}
\rm M_{HI}=\mu m_H \int N_H dA = \mu m_H D^2 \int N_H d\Omega 
\label{eq:M_HI}
\end{equation}
\end{center}

Aperture photometry was done in the N$_{\rm H}$ map of \cite{oosterloo2005} 
to obtain the integrated column densities for our sources.
We used 22\arcsec~apertures to be consistent with our CO(1--0) observations . 
Since these apertures are smaller than the spatial resolution of the HI map (18 $\times$ 95.1 arcsec), 
the photometry results are equivalent to the pixel value of the HI map at the position of our CO targets.
These values are listed in column 3 of Table \ref{tab:S-K_values}.
Then, by dividing Eq. \ref{eq:M_HI} by the CO(1--0) beam solid angle $\rm \Omega=\pi R^2$, we obtain the HI surface densities $\Sigma_{\rm HI}$ listed in column 4 of 
Table \ref{tab:S-K_values}.

Finally, we estimate the SFR surface density $\rm \Sigma_{SFR}$ directly from the H$_{\rm \alpha}$ emission in these regions. From \cite{kennicutt2012}:

\begin{center}
\begin{equation}
\rm log~  SFR[M_{\odot}yr^{-1}]=log~ \rm L_{H_{\alpha}}[erg~ s^{-1}] - 41.27
\label{eq:SFR}
\end{equation}
\end{center}

which, divided by the CO(1--0) beam solid angle $\Omega$, gives the SFR surface density $\Sigma_{\rm SFR}$.
H$_{\rm \alpha}$ luminosities for HaR-1, HaR-2 and HaR-3-4 were obtained from \cite{yagi2013}, and are listed in Table \ref{tab:S-K_values}, along with the corresponding
$\rm \Sigma_{SFR}$. 
For the remaining sources we used the H$_{\alpha}$ map from \cite{kenney2008}, using an aperture photometry of $\sim$1\arcsec~in diameter,
similar to the seeing of the H$_{\alpha}$ observations from \cite{yagi2013}.
For the sources with no visible detection in this map, an upper limit was calculated form the noise level of Kenney's map, estimated in
$\rm 4.8\times10^{-7}erg~s^{-1}cm^{-2}sr^{-1}$ .

\begin{table*}
\caption{ Schmidt-Kennicutt relation values. \label{tab:S-K_values}}
\centering
\begin{tabular}{ccccccc}
\hline\hline
Source &  $\Sigma_{\rm H_2}$  & $\int \rm N_{\rm HI} d\Omega$\tablefootmark{a} & $\Sigma_{\rm HI}$ & log(L$_{\rm H_{ \alpha}})$\tablefootmark{b}& log($\Sigma_{\rm SFR}$) & $\tau_{\rm dep}(\rm H_2)$\\
 &  [M$_{\odot}$pc$^{-2}$] & [10$^{12}$cm$^{-2}$] & [M$_{\odot}$pc$^{-2}$]  & [erg s$^{-1}$] & [M$_{\odot}$yr$^{-1}$kpc$^{-2}$] & [yr]\\
\hline
Source-1 & 0.93 & 0.33 & 0.40 & <35.51 & <-6.19 &  >1.4$\times$10$^{12}$\\
Source-2 & <0.48 & 0.41 & 0.50 & <35.51 & <-6.19 & ---\\
Source-3 & <0.40 & 0.56 & 0.68 & <35.51 & <-6.19 & ---\\
Source-5 & <0.48 & 0.60 & 0.69 & 37.51 & -4.20 & <7.6$\times$10$^{9}$\\
HaR-2  & 0.39 & 2.12  & 2.5 &  37.10 &  -4.61 & 1.6$\times$10$^{10}$\\
HaR-3-4 & <0.40 & --- & --- & 37.75 & -3.96 & <3.6$\times$10$^{9}$\\
HaR-2-1 & <0.44 & 5.56 & 6.77 & <35.51 & <-6.19 & ---\\
HaR-2-2 & <0.44 & 5.13 & 6.24 &<35.51 & <-6.19 & ---\\
HaR-2-3 & <0.22 & 3.67 & 4.46 &<35.51 & <-6.19 & ---\\
HaR-2-4  & 0.24 & 3.08 & 3.74 & <35.51 & <-6.19 & >3.7$\times$10$^{11}$\\
HaR-1 & 0.88  & 2.05  & 2.49 & 35.89  & -5.82 &  5.8$\times$10$^{11}$\\
\hline
\end{tabular}
\tablefoot{Surface densities consider a solid angle $\Omega$=2.74 kpc$^2$, equivalent to the CO(1--0) beamsize at the distance of the source (17.5 Mpc fom \citealt{mei2007}).\\
\tablefoottext{a}{from \cite{oosterloo2005}.}\\
\tablefoottext{b}{from \cite{yagi2013} and \cite{kenney2008}.}
}
\end{table*}

\begin{figure*}
\centering
\includegraphics[width=0.7\textwidth]{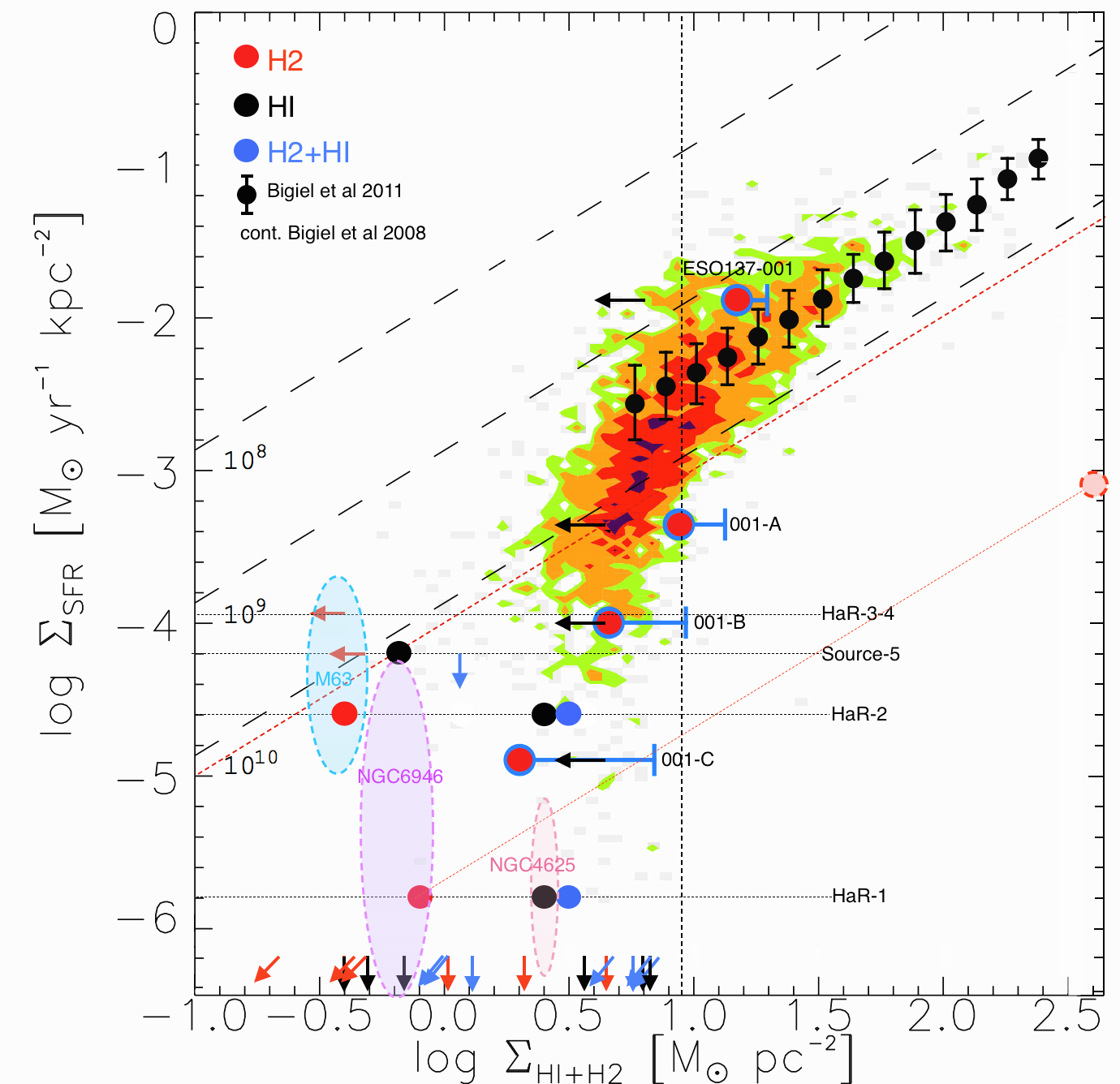}
\caption{S-K relation for sources in Table \ref{tab:S-K_values}, with filled circles and arrows for values and upper limits respectively. 
\label{fig:S-K_relation}
Red markers consider only H$_2$ for $\Sigma_{\rm gas}$, black ones only HI, and blue ones the sum of both.
Figure adapted from \cite{jachym2014}, where their sources ESO137-001, 001-A, 001-B and 001-C have been plotted in a similar way than ours:
red circles for H$_2$ gas, black left arrows for HI upper limits, and blue "error bars" to account for both gas components.
Colored contours account for the spiral galaxies from \cite{bigiel2008}
(green, orange, red and purple for 1, 2, 5 and 10 sampling points per 0.05 dex respectively).
Black markers with error bars correspond to the running medians in $\Sigma_{\rm SFR}$ as a function of $\sigma_{\rm H_2}$ of 30 nearby galaxies from \cite{bigiel2011}.
Shaded ovals represent the data from the outer parts of XUV disk galaxies: NGC4625 and NGC6946 from \cite{watson2014} (priv. communication)  ,
and M63 (NGC5055) from \cite{dessauges2014},
taking only H$_2$ into account in all of them.
The dashed vertical line shows the 9 M$_{\odot}\rm pc^{-2}$ threshold at which the atomic gas saturates.
Dashed inclined lines represent ``isochrones'' of constant star formation efficiencies,
indicating the depletion times $\tau_{\rm dep}=\Sigma_{\rm gas}/\Sigma_{\rm SFR}$ of 10$^8$, 10$^9$ and 10$^{10}$ years to consume all the gas.
The dashed red isochrone marks a depletion time equal to the age of the universe, as one Hubble time (13.8 Gyr). 
A representative shift of the HaR-1 marker for $\Sigma_{\rm H_2}$ is drawn, to show the "effective" molecular gas density
at which stars would formed in this region if we consider a beam correction factor of +2.68 in log space, to convert our 22\arcsec~beam to  a $\sim$1\arcsec~beam as in the H$_{\alpha}$ data.
}
\end{figure*}

The SFR surface densities are plotted as a function of the gas surface densities to construct a  S-K relation in Figure \ref{fig:S-K_relation},
using the values from Table \ref{tab:S-K_values}.
We have plotted separately, the atomic and molecular gas component of $\Sigma_{\rm gas}$, along with the combination of both,
in red, black and blue markers respectively. Arrows denote the upper limits values from Table \ref{tab:S-K_values}.
In this Figure, adapted from \cite{jachym2014}, we can compare our sources
with theirs in the Norma Cluster, as well as with the sample of spiral galaxies from \cite{bigiel2008} 
(in coloured contours), and the sample of 30 nearby galaxies from \cite{bigiel2011}, plotted as the running medians of $\Sigma_{\rm SFR}$ 
as a function of $\Sigma_{\rm H_2}$, and with a typical depletion time  of $\sim$2.3 Gyr. 
Additionally, shaded regions have been included to represent regions from the outskirts of XUV disk galaxies. 
NGC4625 and NGC6946 data was taken from \cite{watson2014} (priv. communication), 
including IRAM-30m CO observations, and H$_{\alpha}$ luminosities measured within a 6\arcsec~aperture.
M63 (NGC5055) data corresponds to the bright UV region located at 1.35r$_{25}$ in \cite{dessauges2014}, with IRAM 30-m CO data,
and a SFR measured from the FUV and 24$\mu$m emission.

``Isochrones'' of constant star formation efficiencies
are also shown to indicate the depletion times of 10$^8$, 10$^9$ and 10$^{10}$ years to consume all the gas,
including an additional red isochrone to mark the age of the universe as one Hubble time (i.e $\tau_{\rm dep}=13.8$Gyr).

Contrary to the photometry done in the HI data, the H$_{\alpha}$ data was not measured in a 22\arcsec~diameter aperture, 
as the CO(1--0) FWHM, but in an aperture of 1\arcsec~in diameter, similar to
the seeing of those observations. 
Since we are averaging this H$_{\alpha}$ emission in a 22\arcsec~aperture  to calculate the $\Sigma_{\rm SFR}$, we could be
diluting the real surface density of the gas being converted into stars. 
To correct for this beam dilution, as a representation we have shifted one of the points in Figure \ref{fig:S-K_relation}
to a fictitious $\Sigma_{\rm H_2}$, corresponding to a source's solid angle of 1\arcsec~in diameter.
This correction translates in a +2.68 shift in log space, and is a representation of the real gas surface density at which
stars would be formed, but always with the same $\tau_{\rm dep}$.

For Source-5 we have made the assumption that the HI and H$\alpha$ emission are spatially correlated, i. e that they both belong
to the gas plume associated to NGC4388. We know that this is true for HI, since we used the data from \cite{oosterloo2005} and
they probed the physical asociation of the HI gas plume to NGC4388. But this could not 
be the case for H$\alpha$, since we used the data from \cite{kenney2008} and their H$\alpha$ map 
show that this source could be associated to M86 instead, when H$\alpha$ is considered.

From Figure \ref{fig:S-K_relation} we can see that our  sources have extremely low SFRs in comparison with the nearby spiral galaxies, 
and are only comparable with the most outer clumps in the H$_{\alpha}$/X-ray tail of the ISM stripped galaxy ESO137-001 in the Norma Cluster
\citep{jachym2014} and the XUV disk galaxies from \cite{watson2014} and \cite{dessauges2014}. 
We obtain depletion times  significantly large. 
For example, HaR-1 and HaR-2 have $\tau_{\rm dep}$ values of  2.2$\times$10$^{12}$ and 1.2$\times$10$^{11}$ years respectively, 
to consume all the amount of gas present (HI+H$_2$).
These values transform into 1.6$\times$10$^{12}$ and 1.0$\times$10$^{11}$ years if we consider only the atomic gas component,
and into  5.8$\times$10$^{11}$ and 1.6$\times$10$^{10}$ years if we consider only the molecular one. Such values are quite large
in comparison with the typical $\tau_{\rm dep}$ of $\sim$2 Gyrs for spiral galaxies, and are even larger than a Hubble time by up to 2 orders of magnitude. 
In Table \ref{tab:S-K_values}, depletion times of H$_2$ that can be calculated are listed in column 7.

The extremely low SF efficiency of our sources seems to fall off the linearity of the S-K relation for typical spiral galaxies at higher gas densities,
a result previously reported in other low gas density environments, such as XUV disk galaxies \citep{dessauges2014}.
\cite{watson2014} presents a different conclusion for their results in XUVs, with a typical SFR in agreement with the S-K linear regime,
but they take into account the 24$\mu$m emission in the SFR, and neglect the contribution of heavy elements in the $\Sigma_{\rm H_2}$.
We see that when we correct by these differences to make their data analytically compatible with ours (i.e neglect the 24$\mu$m
emission and correct for heavy elements), their data points in the S-K plot are comparable to ours.

\section{Summary and Conclusions}

From our molecular cloud and star formation study in the tail north of NGC 4388
in Virgo, we can draw the following conclusions:

\begin{enumerate}
\item CO(1--0) and CO(2--1) observations were done with the IRAM 30-m telescope in a total of 11 targets
all along the ram-pressure stripped tail northeast NGC4388 in the Virgo Cluster, in order to probe the 
presence of molecular gas under extreme conditions. Such targets were selected for having strong peaks of  HI and H${\alpha}$ emission.

\item Four of such positions showed CO detections, and 3 of them concentrated in the HaR-2 region, at a distance of $\sim$70 kpc of NGC4388,
where molecular gas in unexpected.
Given the large distances of these sources to NGC4388, it is not likely that the molecular gas was stripped from the galaxy, 
and must have formed in situ from the HI gas plume.

\item Gaussian line profiles were fitted to the spectra of the detections, finding a range of velocity dispersion
between 12 and 35  km/s. The CO(1--0) line profiles were used to estimate molecular gas masses and surface densities.
The amount of molecular gas in these 3 regions (HaR-1, HaR-2 and HaR-2-4) is very low ( between 0.7 and 2.4 $\times$10$^6$M$_{\odot}$),
and their H$_2$ surface densities between 0.2 and 0.9 M$_{\odot}$pc$^{-2}$. 
These values are well below the HI-H$_2$ threshold, where the gas is mainly atomic and very little is know about the
SFR at such low gas densities, hence the importance of these detections.

\item Using complementary data from \cite{yagi2013} and \cite{kenney2008} for H${\alpha}$ and from \cite{oosterloo2005} for HI, we
computed $\Sigma_{\rm SFR}$ and $\Sigma_{\rm HI}$ to plot , in combination with $\Sigma_{\rm H_2}$, a S-K relation.
Our sources show an extremely low SFR (up to 2 order of magnitude lower than for typical spiral galaxies).
For example, HaR-1 and HaR-2 have total gas depletion times of 2.2$\times$10$^{12}$ and 1.2$\times$10$^{11}$ years respectively.
If we consider just the molecular gas component, these depletion times are 5.8$\times$10$^{11}$ and 1.6$\times$10$^{10}$ years.
Furthermore, Source-1 and HaR-2-4 have H$_2$ depletion times even greater than 1.4 $\times$10$^{12}$ and 3.7 $\times$10$^{11}$ years respectively.
These high values of depletion times exceed by far a Hubble time, thus indicating that this molecular gas will not form stars eventually, but remain 
in a gaseous phase and later join the ICM.

\item From Figure \ref{fig:S-K_relation} we can see the linearity between the SFR and the gas surface density, so clear at
high gas surface densities (> 9 M$_{\odot}$pc$^{-2}$) for normal spiral galaxies, cannot be extrapolated to lower densities,
below the HI-H$_2$ threshold, where the star formation is extremely inefficient, and the molecular gas, even though present, does not
necessarily form stars .

\end{enumerate}
\begin{acknowledgements}
We warmly thank the referee for constructive comments and suggestions.
Also, the IRAM staff is gratefully acknowledged for their
help in the data acquisition.
We thank T. Oosterloo and J. Kenney for facilitating important HI and H$_{\alpha}$ data, respectively. 
F.C. acknowledges the European Research Council
for the Advanced Grant Program Number 267399-Momentum.
We made use of the NASA/IPAC Extragalactic Database (NED),
and of the HyperLeda database.
C.V. acknowledges financial support from CNRS and CONICYT 
through agreement signed on December 11th 2007.

\end{acknowledgements}


\end{document}